\documentclass[preprint,12pt]{elsarticle}

\usepackage{graphicx}
\usepackage{amssymb}

\journal{Surface Science}

\begin{document}

\begin{frontmatter}


\title{Sputtering of Oxygen Ice by Low Energy Ions}



\author[1]{E.~A.~Muntean}
\ead{emuntean01@qub.ac.uk}
\author[2]{P.~Lacerda\fnref{vrf}}
\author[1]{T.~A.~Field}
\author[1]{A.~Fitzsimmons}
\author[]{C.~A.~Hunniford\fnref{vrf}}
\author[1]{R.~W.~McCullough}
\address[1]{Department of Physics and Astronomy, School of Mathematics and Physics, Queen's University Belfast, BT7 1NN, N. Ireland, UK}
\address[2]{Planck Institute for Solar System Research, G\"ottingen, 37191, Germany}
\fntext[vrf]{Visiting Researcher at Queen's University Belfast}

\begin{abstract}	
Naturally occurring ices lie on both interstellar dust grains and on celestial objects, such as those in the outer solar system. These ices are continuously subjected to irradiation by ions from the solar wind and/or cosmic rays, which modify their surfaces. As a result, new molecular species may form which can be sputtered off into space or planetary atmospheres. We determined the experimental values of sputtering yields for irradiation of oxygen ice at 10 K by singly (He$^+$, C$^+$, N$^+$, O$^+$ and Ar$^+$) and doubly (C$^{2+}$, N$^{2+}$ and O$^{2+}$) charged ions with 4 keV kinetic energy. In these laboratory experiments, oxygen ice was deposited and irradiated by ions in an ultra high vacuum chamber at low temperature to simulate the environment of space. The number of molecules removed by sputtering was observed by measurement of the ice thickness using laser interferometry. Preliminary mass spectra were taken of sputtered species and of molecules formed in the ice by temperature programmed desorption (TPD). We find that the experimental sputtering yields increase approximately linearly with the projectile ion mass (or momentum squared) for all ions studied. No difference was found between the sputtering yield for singly and doubly charged ions of the same atom within the experimental uncertainty, as expected for a process dominated by momentum transfer. The experimental sputter yields are in good agreement with values calculated using a theoretical model except in the case of oxygen ions. Preliminary studies have shown molecular oxygen as the dominant species sputtered and TPD measurements indicate ozone formation.
\end{abstract}


\end{frontmatter}


\section{Introduction}
\label{S:1}

Studies of ion-induced processing of astrophysical ice analogues in the laboratory are relevant to a variety of different environments, such as the icy mantles of dust grains in the interstellar medium and protoplanetary discs, together with the surfaces of Solar system objects such as comets, Centaurs, Kuiper belt objects, and the icy satellites of the outer planets. Sputtering by ion impact can give rise to changes in the chemical, physical and optical properties of the ice as a result of both elastic and inelastic collisions. In the sputtering process, material is ejected from the surface which may assist in the development of thin atmospheres such as those found around Pluto and the icy moons of Jupiter and Saturn. An extreme example is Europa, where O$_2$ on the trailing hemisphere of Jupiter's moon Europa is sputtered by low-energy ions to become the dominant component of the moon's atmosphere and neutral gas torus \cite{Cassidy,Johnson,Hall}, together with causing the hemispherical colour variations.

Sputtering also plays a role in the chemical alteration of distant comet surfaces, and contributes to the formation of the first coma as these objects approach the sun. Recently, the signature of this process was recorded by the ROSINA mass spectrometers on-board the Rosetta spacecraft at comet 67P/Churyumov-Gerasimenko. Small amounts of refractory elements such as Na and Si are believed to have been generated by Solar wind sputtering of dust grains on the surface of the nucleus \cite{Wurz}. 

O$_2$ ice is not a major component of the observed astrophysical ices inventory but O$_2$ ice processing by the slow component of the Solar wind and cosmic rays plays an important role in the formation of other species such as ozone \cite{Ennis}, water and various carbon-oxides \cite{Jing}.  Whilst there have been many sputtering studies using water ice as a target; (see \cite{Fama} and references within) sputtering of oxygen ices has seen fewer investigations. Given that cosmic ray impacts onto molecular oxygen on dust grain surfaces is one possible pathway in the formation of water (e.g., \cite{Jing}), the associated loss pathway of sputtering warrants further investigation. Depending on the energy of the incoming projectile ion, experimental sputtering can be divided into two main categories: sputtering by low energy ions (a few keV) and sputtering by high energy ions ($>$30 keV, up to $\sim$3 MeV).  At energies of a few keV ions primary lose energy in `nuclear' ballistic collisions, where target molecules are excited vibrationally, rotationally and translationally. Above 30 keV energy is lost primarily in electronic excitation of target molecules by, e.g., ionisation. In between these two energy ranges neither of these two energy loss mechanisms dominates.

Sputtering of oxygen ice by low energy (4-10 keV) H$^+$, H$_2^+$, and H$_3^+$ and  high energy ions He$^+$ and H$^+$ (up to 3.5 MeV) has been intensively studied by \citet{Ellegaard94,Ellegaard86}, and references within.  They concluded that oxygen ice sputters more efficiently than nitrogen ice by a factor of almost 2, and that the sputter yields are proportional to the electronic stopping power, i.e. the energy lost by the projectile ion due to inelastic collisions with bound electrons as it penetrates the ice. The effect of multiply-charged ions on ion-induced sputtering of O$_2$ has been neglected so far in the literature, with the only study at low energy for oxygen ices by \citet{Gibbs} on oxygen ice irradiated at MeV energies by He$^{2+}$ and He$^{+}$ ions. They concluded that the yields are a factor of 1.2 higher for He$^{2+}$ ions. 

The present work is a laboratory study of sputtering of oxygen ice by 4 keV singly and doubly charged C, N, and O ions and singly charged He and Ar. There have been only a few laboratory experiments on sputtering by low energy non-noble gas ionic species such as C, N and O, which have the potential to be incorporated in new molecules in surfaces and even fewer experiments for sputtering by multiply charged ions.

\begin{figure}[ht]
\centering
\includegraphics[width=0.61\columnwidth]{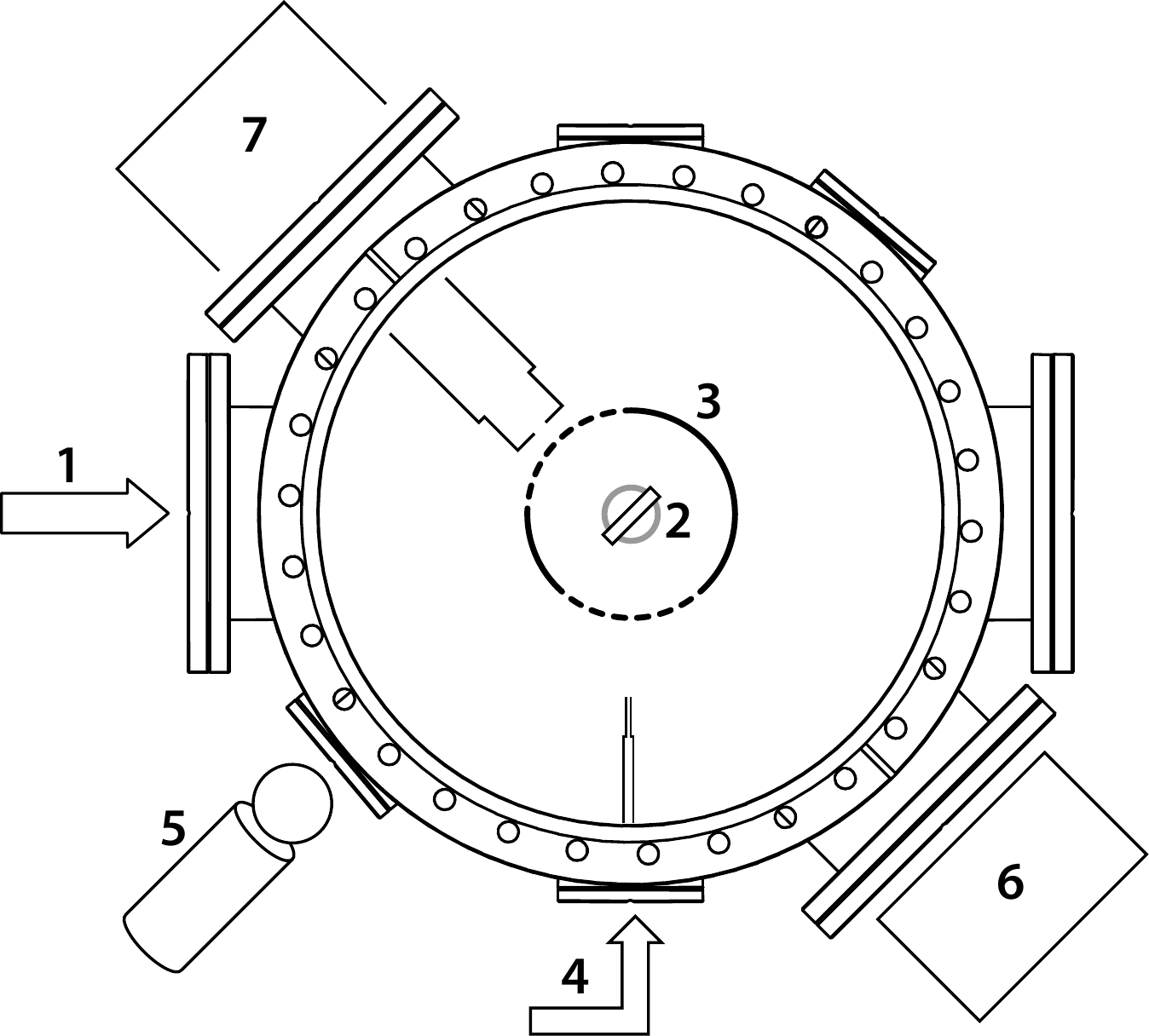}
\caption{Schematic diagram of the experimental arrangement with 1) ion irradiation part, 2) KBr substrate, 3) radiation shield, 4) vapour deposition nozzle , 5) and 6) thickness monitor  comprising (5) a 405 nm diode laser and (6) a photodiode detector and 7) line of sight quadrupole mass spectrometer.}
\label{fig:Chamber}
\end{figure}

\section{Experimental details}
\label{S:2}

The experimental arrangement used in the present work is shown schematically in Figure \ref{fig:Chamber}.  An Ultra High Vacuum (UHV) chamber was coupled to a low energy ion accelerator \cite{Broetz} equipped with a 10 GHz Electron Cyclotron Resonance (ECR) ion source \cite{Hunniford}.  Singly and doubly charged ions of $^{13}$C, $^{14}$N, $^{16}$O and singly charged ions of $^4$He  and $^{40}$Ar, at 4 keV, were focused and directed into the experimental chamber (which operated at a base pressure $\sim 1 \times 10^{-9}$ mbar).  Ion beams were collimated by three 4 mm diameter apertures separated by 50, 38 and 26 mm respectively from the substrate providing an ion beam of 4 mm in diameter with less than 1\% increase in diameter for the maximum current density used at the ice surface.  A potassium bromide (KBr) substrate, 20 mm in diameter and 2 mm thick, was clamped to an earthed, temperature controlled sample holder (A.S.\ Scientific) surrounded by a gold plated copper radiation shield.

The substrate temperature was controlled to within 0.1 K at temperatures from 8 to 300 K using a calibrated silicon diode.  A slot in the radiation shield 15 mm high and with a total angular width of 120 degrees enables ion beam access to the deposited ices; the slot gives access of $\pm$60 degrees with respect to the normal of the substrate surface in the radiation shield.  The radiation shield temperature was measured throughout the experiment by a diode, and was found to be constant at a value of 48 K during the irradiation and heating procedure. The cold-head was rotatable through 360 degrees via a differentially pumped rotary stage.  The substrate could be rotated to any desired position during the experiment. Additionally, as shown schematically in Figure \ref{fig:Chamber} a quadrupole mass spectrometer (QMS; model HAL-201 from Hiden Analytical) could be used to monitor species emanating from the sample either from sputtering or thermal desorption.

\begin{figure}
\centering
\includegraphics[width=0.61\columnwidth]{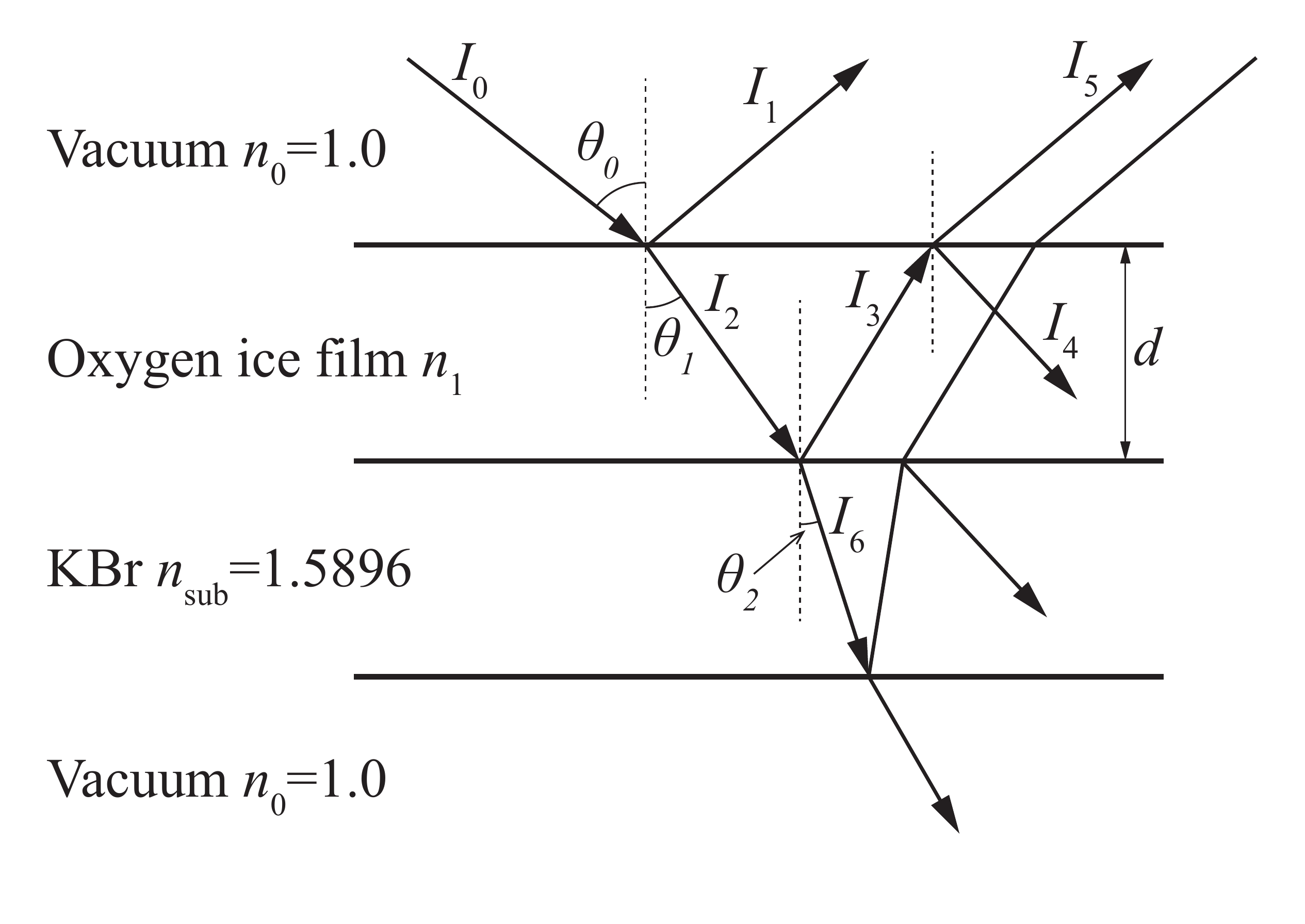}
\caption{Reflection and refraction model for an oxygen ice film deposition on a KBr window.}
\label{fig:Reflection}
\end{figure}

The ice films were deposited at a pressure of $1.6\times 10^{-7}$ mbar with the thickness measured by laser interferometry using a 405 nm diode laser and a photodiode detector.  During deposition, the cold-head assembly was rotated such that the vapour deposition nozzle was directed normal to the substrate surface, at a distance of 25 mm, with the laser and detector at 45 degrees with respect to the substrate normal.  Oxygen ice films of up to 258 nm were deposited onto the substrate which was held at 10 K.
 
Figure \ref{fig:Reflection} shows a schematic diagram of the reflection and refraction for the oxygen ice film deposited on a KBr window.  $\theta_0$ is 45 degrees and  represents the angle of incidence of the laser light with respect to the surface normal, and $\theta_1$ is the angle of refraction at the vacuum ice interface calculated from Snell’s Law (\ref{eq:snellslaw}). $n_0$ is the vacuum refractive index and $n_1$ the oxygen ice refractive index.
\begin{equation}
n_0 \sin\theta_0 = n_1 \sin\theta_1.
\label{eq:snellslaw}
\end{equation}

\noindent The intensities $I_1$ to $I_5$ can be calculated from the reflection and refraction laws. The model fit equation is expressed as 
\begin{equation}
\mathrm{PS} = C \left( I_1 + T^{2 d/\cos\theta_1} \times I_5 \cos\frac{\pi d}{\cos\theta_1}\right)
\label{eq:photodiodesignal}
\end{equation}

\noindent where PS is the photodiode signal in volts, $I_1$ is the reflection intensity from vacuum to the ice and $I_5$ the refracted intensity after the light passes through the ice and is reflected back. In the model a value 1.285 for the refractive index of the ice gives the best fit. $C$ is a constant related to the photodiode efficiency and $T$ is the transmission coefficient of the ice at 405 nm.  In equation (\ref{eq:d}), $m$ is an integer, where odd values of $m$ correspond to maxima (constructive interference) and even values of $m$ correspond to minima (destructive interference) at intervals of 86 nm for this wavelength.  

\begin{equation}
d=m\frac{\lambda}{4}\cos\theta_1
\label{eq:d}
\end{equation}

The ice density was calculated using the Lorentz-Lorenz relation \cite{Fulvio}
\begin{equation}
L\times\rho=\frac{n^2-1}{n^2+2}
\label{eq:lorentz}
\end{equation}

\noindent where $L$ is the Lorentz-Lorenz factor, $n$ is the refractive index of the ice and $\rho$ the density of the ice. A value of $L=0.1294$ cm$^3$ g$^{-1}$ has been used in the present work \cite{Fulvio} and we assume that $L$ is constant in the visible wave range. Using equation (\ref{eq:lorentz}) an oxygen ice density of $\rho=1.38\pm0.20$ g cm$^{-3}$ was obtained.

\begin{figure}[ht]
\centering
\includegraphics[width=0.7\columnwidth]{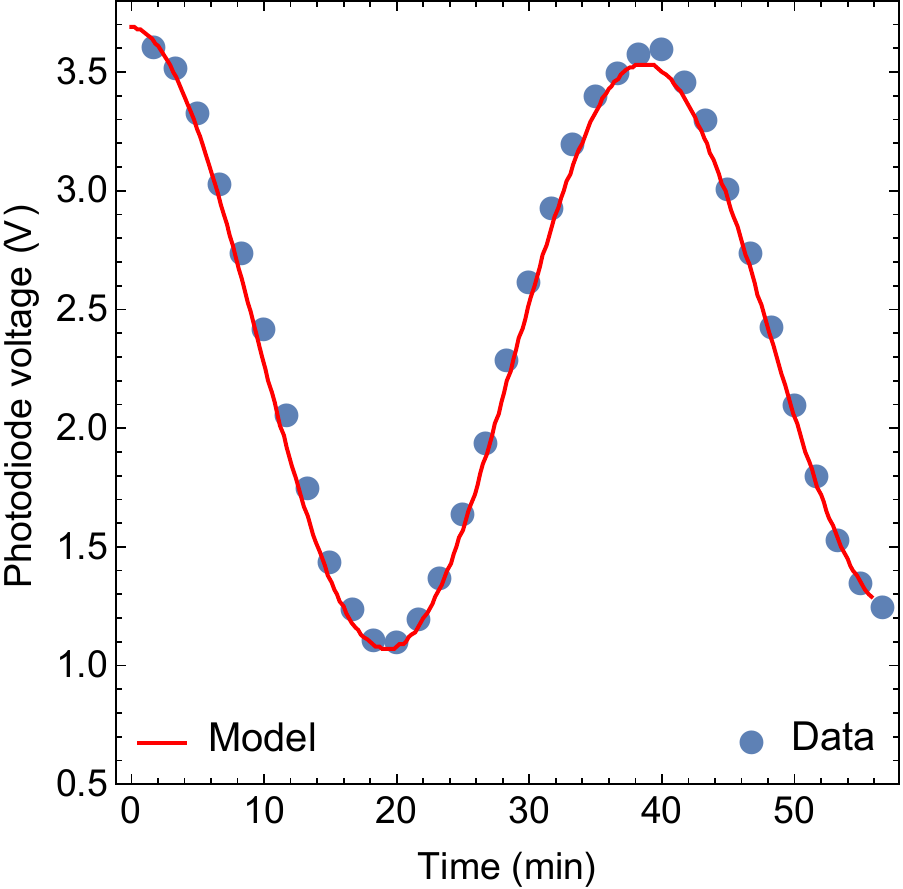}
\caption{Laser interferometer photodiode voltage (points) for experimental values and model fit (solid line) vs.\ time during the formation of oxygen ice at 10 K from oxygen gas at a pressure of $1.6 \times 10^{-7}$ mbar.}
\label{fig:DepositionO}
\end{figure}

\citet{Fulvio} measured an oxygen ice refractive index of $n=1.322$ by laser interferometry and used this value to obtain an oxygen ice density of 1.54 g cm$^{-3}$ at 16 K, again using a Lorentz-Lorenz value of $L= 0.1294$ cm$^3$ g$^{-1}$.  \citet{Roux} found a refractive index of 1.25 for oxygen ice at 20 K. Therefore our measurements are in line with previous determinations of refractive index and density.

\begin{figure}[ht]
\centering
\includegraphics[width=0.7\columnwidth]{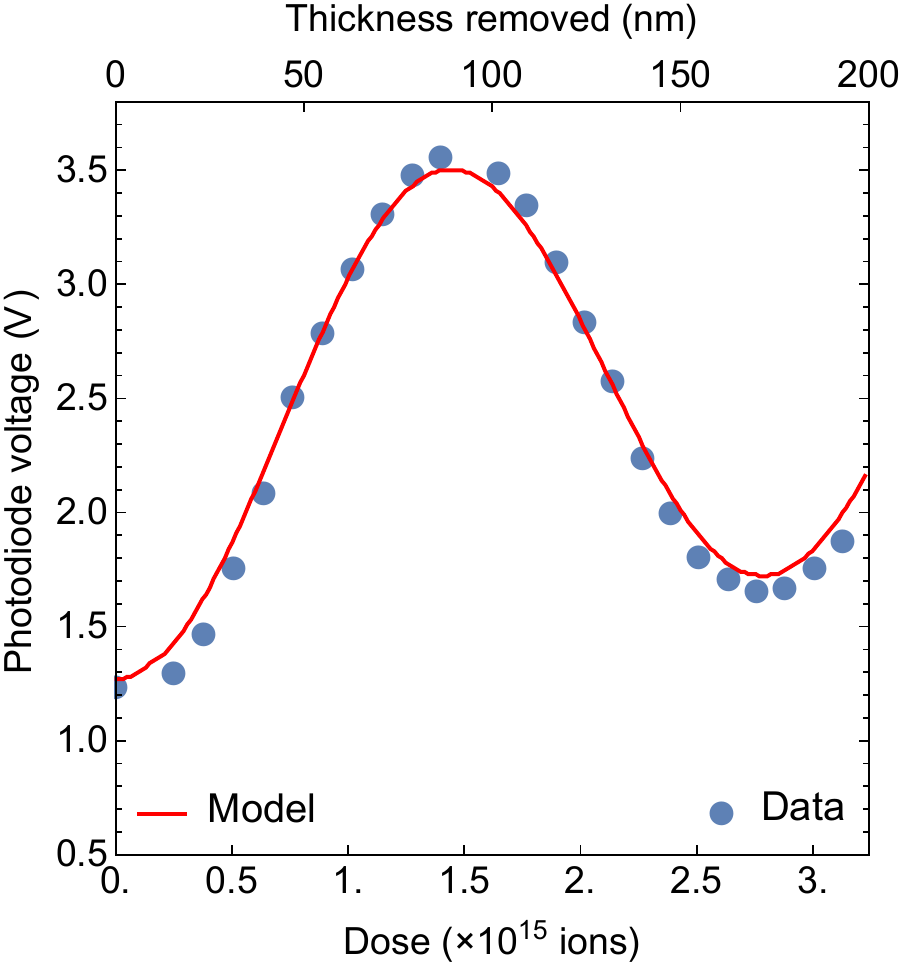}
\caption{Laser interferometric data (points) and sputtering model (solid line) for 4 keV N$^+$ on a 258.4 nm oxygen ice film deposited and irradiated at 10 K.}
\label{fig:DepositionN}
\end{figure}

After a 258 nm film of O$_2$ ice was deposited, the cold-head was rotated to enable irradiation by the ion beam, at 45 degrees with respect to the normal to the ice film surface. The ion beam current was measured by a Faraday cup inserted immediately in front of the ice film and the current was monitored during irradiation of the ice sample by a 90\% transmission metal mesh. Both the Faraday cup and the mesh were positively biased at 50 V to suppress secondary electron emission. The ice film thickness was measured by the laser interferometer after a specific ion irradiation dose and this process repeated for further ion doses. 

 Figure \ref{fig:DepositionO} shows the interference fringes from the photodiode together with a fit from the model as described above.
Measurements of the reduction of the oxygen ice film thickness by the laser interferometer enabled the sputtering yield $Y_S$ (molecules / incident ion) to be calculated. The ion beam was calculated to be an ellipsoid of area $17.7 \times 10^{-2}$ cm$^2$ giving an ion beam current density  of $\sim0.15$ $\mu$A cm$^{-2}$

Figure \ref{fig:DepositionN} shows the photodiode detector voltage as a function of ion dose for irradiation of a 258 nm oxygen ice film, at 10 K by 4 keV N$^+$ ions. Also shown in Figure \ref{fig:DepositionN} is a model fit (solid line) using the values of refractive index and density determined from the deposition profile shown in Figure \ref{fig:DepositionO}, which allows the determination of the ice thickness removed (top axis).

\begin{figure}[ht]
\centering
\includegraphics[width=0.7\columnwidth]{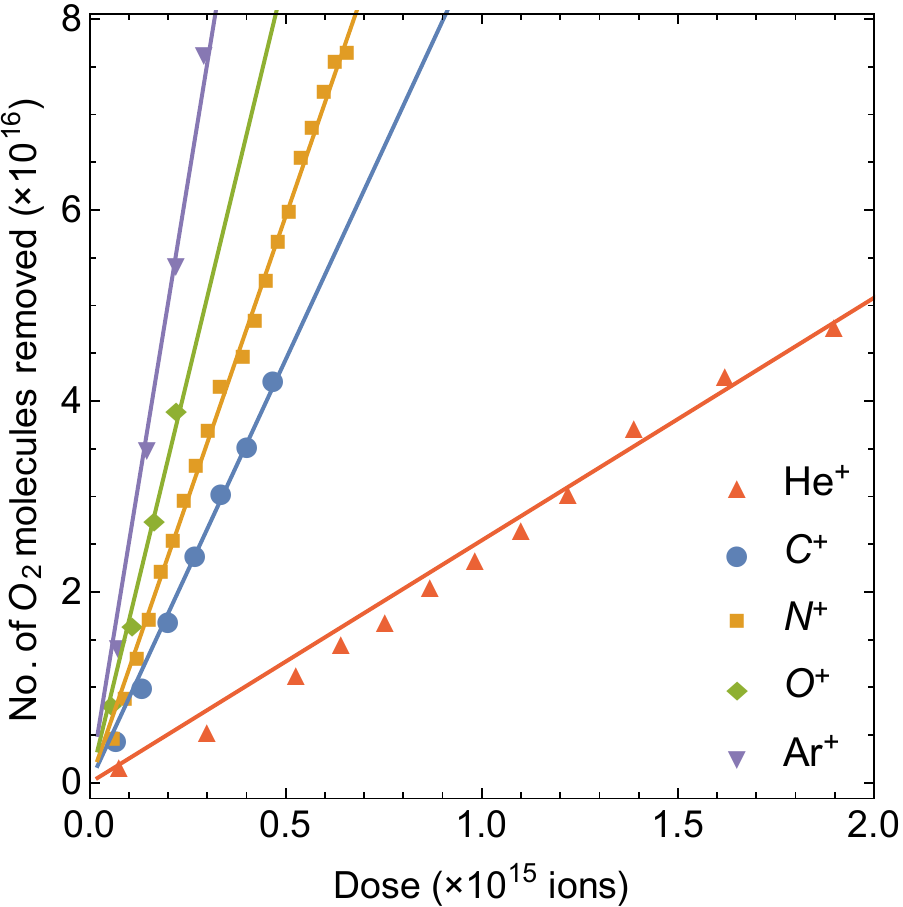}
\caption{Number of O$_2$ molecules removed by 4 keV singly charged ions on oxygen ice at 10 K as a function of ion dose, where upward triangles correspond to $^4$He$^+$, circles correspond to $^{12}$C$^+$, squares correspond to $^{14}$N$^+$, diamonds correspond to $^{16}$O$^+$ and downward triangles correspond to $^{40}$Ar$^+$.}
\label{fig:SinglySputtering}
\end{figure}

\begin{figure}[ht]
\centering
\includegraphics[width=0.7\columnwidth]{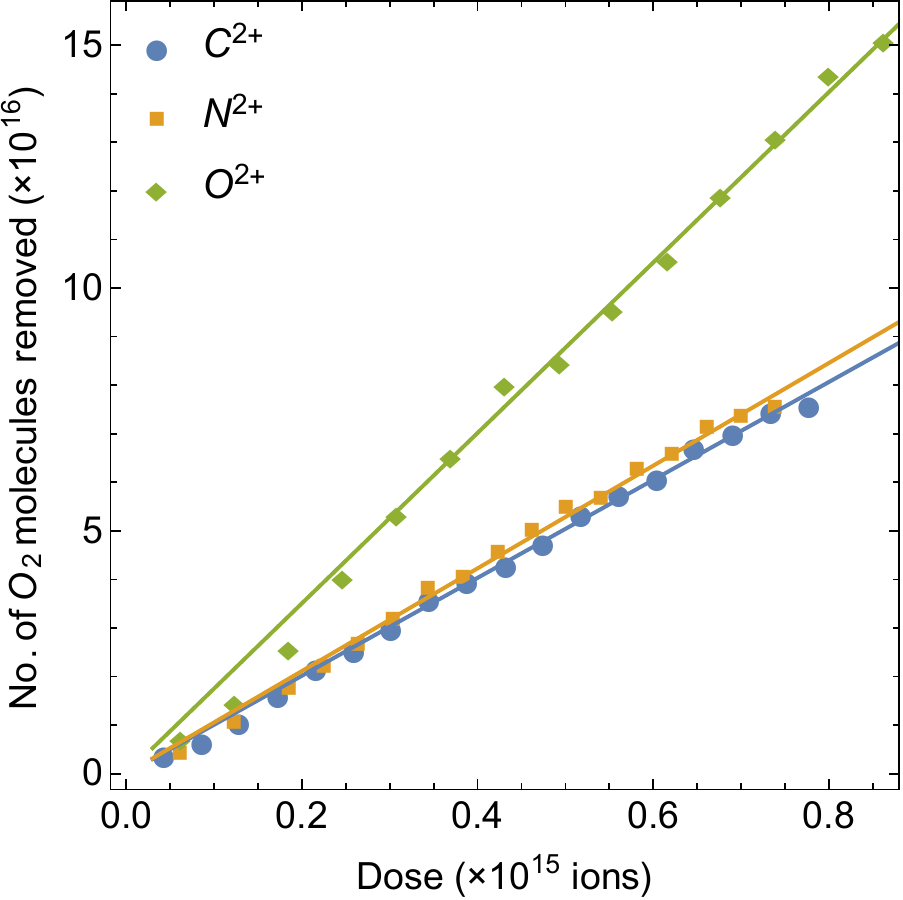}
\caption{Number of O$_2$ molecules removed by 4 keV singly charged ions on oxygen ice at 10 K as a function of ion dose, where circles correspond to $^{12}$C$^{2+}$, squares correspond to $^{14}$N$^{2+}$ and diamonds correspond to $^{16}$O$^{2+}$.}
\label{fig:DoublySputtering}
\end{figure}

\section{Results and discussions}
\label{S:3}

Figure \ref{fig:SinglySputtering} shows the number of  oxygen ice molecules eroded at 10 K, as a function of the dose of $^4$He$^+$, $^{12}$C$^+$, $^{14}$N$^+$, $^{16}$O$^+$ and $^{40}$Ar$^+$ ions, all at 4 keV kinetic energy.  Figure \ref{fig:DoublySputtering} is the same, but for doubly charged $^{12}$C$^{2+}$, $^{14}$N$^{2+}$ and $^{16}$O$^{2+}$ ions.  Note that care was taken to correct the ion acceleration voltage for the charge state of the ion to maintain a constant 4 keV kinetic energy; acceleration of a doubly charged ion through a potential difference of 2 kV gives an increase of 4 keV in kinetic energy whereas a potential difference of 4 kV is required for singly charged ions. It is clear from both Figures \ref{fig:SinglySputtering} and \ref{fig:DoublySputtering} that the number of molecules removed varies approximately linearly with the ion dose in all cases.A linear sputter rate was observed for all studies reported here. The persistence of this linear relationship shows that for all ion exposures there remained sufficient O$_2$ ice to prevent interactions between the ion beam and the KBr substrate. Furthermore, the linearity implies that there was no charging effect of the target ice by the ion beam, which would have reduced the intensity of the impinging ions. The experimental error in the number of molecules removed at each data point in Figures \ref{fig:SinglySputtering} and \ref{fig:DoublySputtering} is estimated to be 16\% from fitting the experimental measurements of light intensity to the model of the light interference patterns.  The gradients of the straight line linear regression fits correspond to sputtering yields, $Y_S$; the number of molecules ejected per incident projectile ion. Experimental sputtering yields resulting from these data are shown in Table 1 with uncertainties reflecting only the random error in determining the gradients of best fit lines to the data in Figures \ref{fig:SinglySputtering} and \ref{fig:DoublySputtering}. As mentioned above, in addition to these random errors there is a potential systematic error of 16\% associated with the determination of the number of molecules removed from the optical measurements. These systematic errors are due to uncertainty in the exact refractive index of the oxygen ice and are, therefore, the same for all the experiments. Thus, experimental data can be reliably compared from one ion to another by taking into consideration only the random errors shown in Table 1, but the additional systematic error should be included when the absolute values for sputtering yields are used in models.  In tests with variable ion beam intensity  the sputtering yields were found to be independent of the ion fluence.

\begin{table}
\caption{Experimental and theoretical values for sputtering yield of oxygen ice at 10 K by 4 keV singly and doubly charged ions. The experimental errors shown are random error in fitting the data shown in Figures \ref{fig:SinglySputtering} and \ref{fig:DoublySputtering}. There may be an additional systematic error estimated to be $\pm$16\% (see text).} 
\label{table:1} 
\centering 
\begin{tabular}{ccccc} 
          & $n$  & $\rho$ (g cm$^{-3}$) &  $Y_S$(O$_2$) Expt. & $Y_S$(O$_2$) Theory \\
          \hline
 He$^+$   & 1.285 & 1.378 & $25\pm1$   & 25 \\
 C$^+$    & " & " & $89\pm5$   & 94 \\
 N$^+$    & " & " & $119\pm7$  & 112 \\
 O$^+$    & " & " & $170\pm10$ & 124 \\
 Ar$^+$   & " & " & $252\pm15$ & 243 \\
 \hline 
 C$^{2+}$ & " & " & $101\pm6$  & $\cdots$ \\
 N$^{2+}$ & " & " & $106\pm6$  & $\cdots$ \\
 O$^{2+}$ & " & " & $176\pm11$ & $\cdots$ \\
\hline 
\end{tabular}
\end{table}

The experimental sputter yields have been compared with values predicted by a model developed by \citet{Fama}, which are also shown in Table 1. Fam\'a's model was originally used to predict the sputtering yield, $Y(E, m_1, Z_1, \theta, T)$, following ion impact on water ice at temperatures, $T$, from 10 K to 140 K with ion energies, $E$, up to 100 keV and projectile ions with mass $m_1$, atomic number $Z_1$ and incident angle $\theta$; their expression is

\begin{equation}
\label {sY:o}
Y\left( E,m_1,Z_1,\theta,T \right) = \\ \frac{1}{U_0}\left(\frac{3}{4\,\pi^2\,C_0}\alpha S_n + \eta\,S_e^2 \right) \times \left( 1 + \frac{Y_1}{Y_0}\exp{\frac{-E_0}{k\,T}} \right)\,\cos^{-f}\theta.
\end{equation}

\begin{figure}[ht]
\centering
\includegraphics[width=0.7\columnwidth]{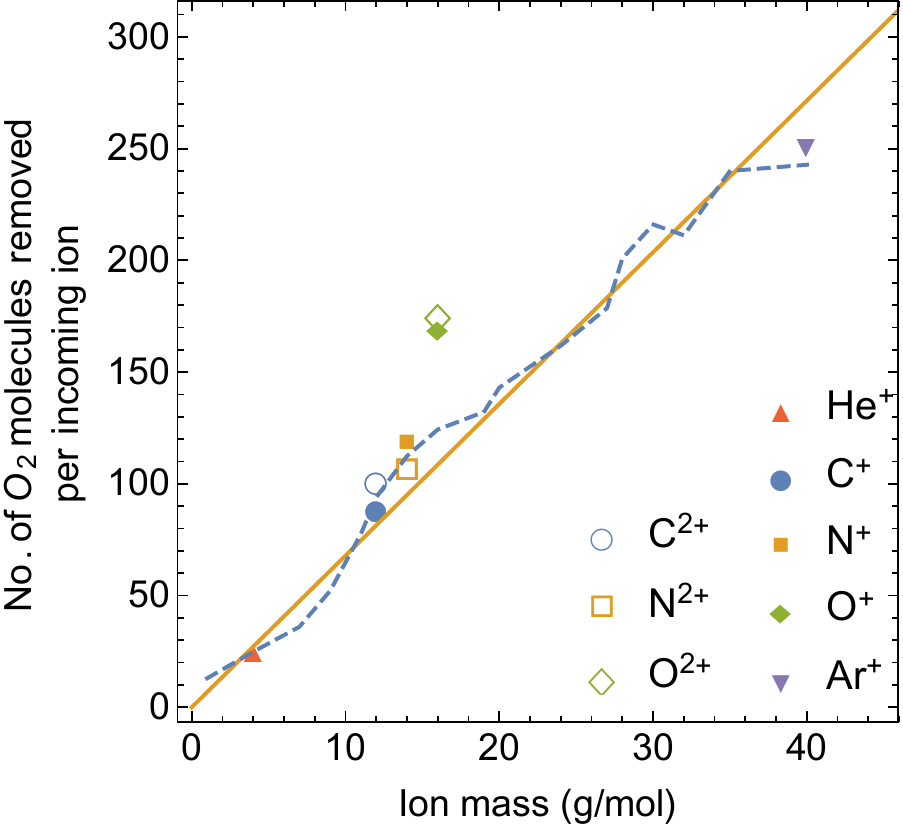}
\caption{Sputtering yield per incoming ion as a function of ion mass. Solid symbols correspond to singly ionised species ($^4$He$^+ $ = upward triangle, $^{12}$C$^+$ = circles, $^{14}$N$^+$ = squares, $^{16}$O$^+$ = diamonds, $^{40}$Ar$^+$ = downward triangles) and empty symbols correspond to doubly ionised species ($^{12}$C$^{2+}$ = circles, $^{14}$N$^{2+}$ = squares, $^{16}$O$^{2+}$ = diamonds.) The dashed line indicates the theoretical yield as a function of projectile mass, and the solid line shows a simple linear fit (slope 6.79) to all experimental data except the oxygen ion values.}
\label{fig:YieldPerIon}
\end{figure}

\noindent Here, $U_0$ is the sublimation energy per molecule.  A value of 0.07 eV/molecule was used in the current work, which is close to the value of 0.09 eV/molecule used by \citet{Ellegaard94,Ellegaard86}.  This value of 0.07 eV/molecule was chosen to maximise the agreement between sputter yields predicted by the model and the experimental values.  It has been reported that the exact value of $U_0$ depends on the deposition conditions and ice thickness \cite{Bienfait}, which may account for the difference between the present value of $U_0$ and the value of Ellegard.  $S_n$ is the nuclear stopping cross-section and $S_e$ is the electronic stopping cross-section; these stopping ranges have been calculated with the Stopping Range of Ions in Matter (SRIM\footnote{http://www.srim.org/}) Programme.  In this work, the energy independent formulae giving values of $\eta$ (a function of the atomic number, $Z_1$, of the projectile), $f$ (the order of the angular dependence, function of the projectile atomic mass, $m_1$) and $\alpha$ (function of the mass ratio of target and projectile, $m_2/m_1$) have been calculated using the empirical formulae given by \citet{Fama}, which are not reproduced here. The formula used by \citet{Fama} to calculate the value of $\eta$ was developed specifically for water ice; it is based on a double exponential fit to a number of previously measured values taken from \citet{Sigmund}. We experimented with the fit, which can be done for any particular mass ratio of target ice molecule to incoming ion, and found very small differences ($<$8\%) between using oxygen ice and water ice, so we decided to retain the original formalism by \citet{Fama}. Note that we use atomic mass 16 rather than molecular mass 18 in the model as we assume that the primary interaction is between the projectile ions and the individual oxygen atoms in the ice. The projectile ion energy, 4 keV, is nearly three orders of magnitude greater than the bond dissociation energy of molecular oxygen, $\sim$5 eV. Thus, although collisions will be between the projectile ion and O$_2$ molecules of mass 32 the bond between the two oxygen atoms is not strong enough to hold the molecule together in collisions where one of the oxygen atoms is struck by the projectile. Furthermore, the stopping ranges calculated using SRIM also consider atomic rather than molecular masses. $C_0$ is a  constant, which is calculated from the differential cross section for elastic scattering in the binary collision approximation \cite{Fama}; here we have determined a value of 1.808 \AA$^2$ from \citet{Sigmund}.  In the present work the temperature, $T$, was constant at 10 K, the energy, $E$, was 4 keV and the angle of incidence, $\theta$, was 45 degrees.  The term $\frac{Y_1}{Y_0}\exp{\frac{-E_0}{k\,T}}$ is neglected here as our experiment was carried out at constant temperature and energy (for details, see \cite{Fama}.)

\begin{figure}[ht]
\centering
\includegraphics[width=0.7\columnwidth]{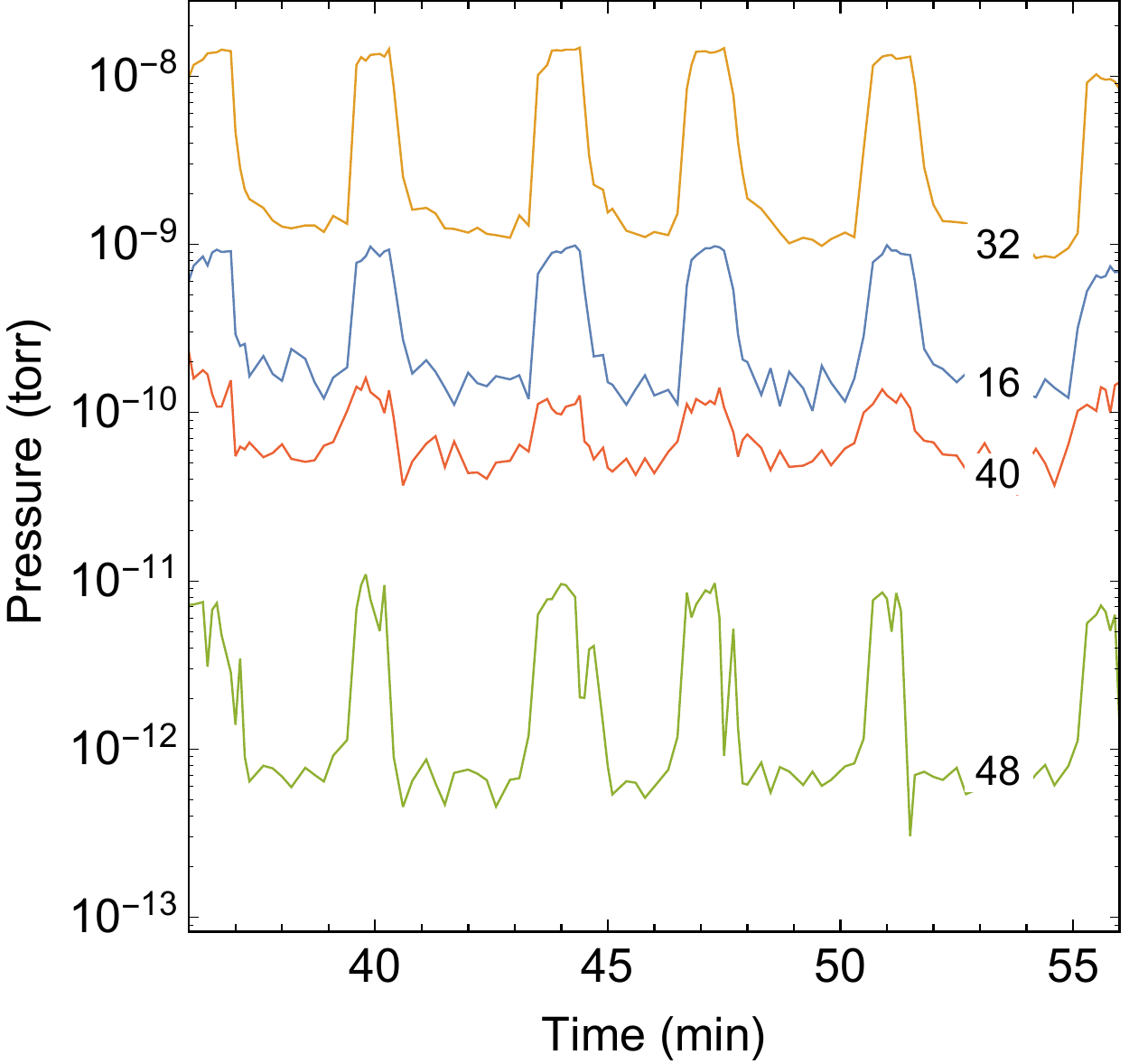}
\caption{Sputtered species due to irradiation of O$_2$ ice by Ar$^+$ ions. Lines indicate pressure in the chamber due to different species labeled according to their mass (mass 16 = O, mass 32 = O$_2$, mass 40 = Ar, mass 48 = O$_3$). The dominant oscillation corresponds to the ion beam being turned on and off.}
\label{fig:O2Sputtering}
\end{figure}

\begin{figure}[ht]
\centering
\includegraphics[width=0.7\columnwidth]{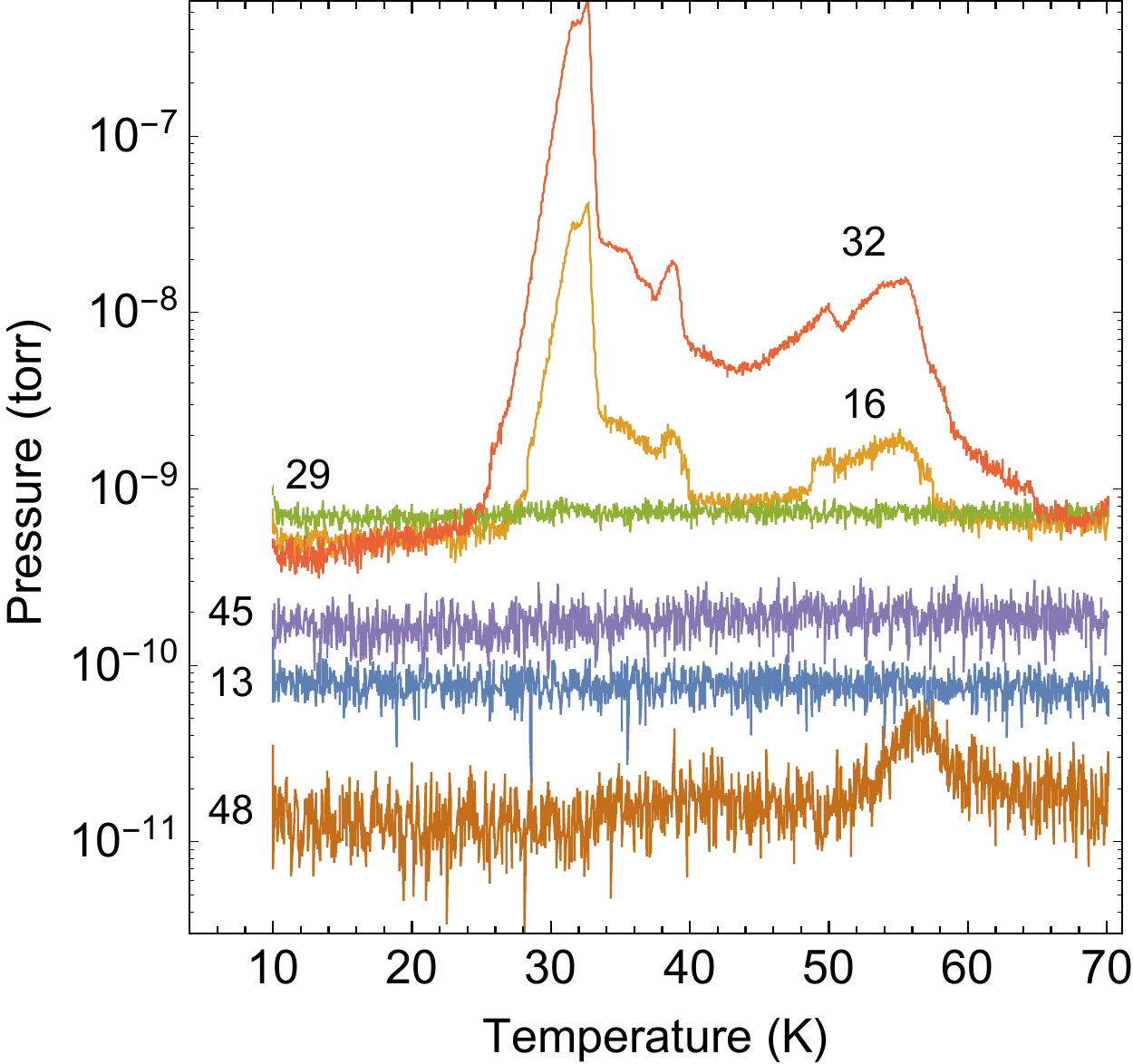}
\caption{Temperature programmed desorption of O$_2$ ice following irradiation by C$^+$ ions. Lines indicate pressure in the chamber due to different species labeled by their mass (mass 13 = $^{13}$C, mass 16 = O, mass 29 = $^{13}$CO, mass 32 = O$_2$, mass 45 = $^{13}$CO$_2$, mass 48 = O$_3$).}
\label{fig:TPD}
\end{figure}

Figure \ref{fig:YieldPerIon} shows our experimental O$_2$ sputtering yields plotted against incoming ion mass. Overplotted are the theoretical prediction using the model just described, and a simple linear fit with a slope 6.79 molecules per unit ion mass to the experimental data. The sputtering yield is therefore proportional to mass of the incoming ion or the square of the momentum. As can be seen in Table \ref{table:1} and Figure \ref{fig:YieldPerIon}, the experimental and theoretically predicted sputtering yields agree with each other remarkably well, particularly considering the model was developed for water ice. The only exception is the case of oxygen ion sputtering, where we measure a $\sim$45\% higher sputtering yield than predicted by theory. We note that the effect is seen in different experimental runs using both singly and doubly charged oxygen ions, which give similar sputtering yields within the experimental error. It is unclear why oxygen ion sputtering yields should be higher than predicted by the model; the matter deserves further investigation.  As this effect is observed only in the case where the target and projectile species are the same, it would be interesting to see whether a similar enhancement is observed in N$_2$ ice irradiated by N ions.

As can be seen in Figure \ref{fig:YieldPerIon}, the sputtering yields increase as the ion momentum increases for both singly and doubly charged ions, showing that momentum transfer alone is the dominant factor for almost all the species studied. In fact, with the exception of the oxygen ions discussed above, a linear relation (6.79 O$_2$ molecules removed per incoming ion per unit mass of incoming ion; solid line in Figure \ref{fig:YieldPerIon}) to our experimental data provides a useful approximation to the more detailed model in the range of ion masses considered. The dashed line in Figure \ref{fig:YieldPerIon} connects sputtering yields calculated with the theoretical model for the highest abundance isotopes of atoms with atomic numbers from 1 to 18; H to Ar.  Electronic stopping dominates in the case of He$^+$ whereas nuclear stopping becomes more important for the heavier ions. Our result can be compared with \citet{Ellegaard94} who studied sputtering of oxygen ice by H$^+$, H$_2^+$ and H$_3^{+}$ with energies from 1-5 keV. We measure a sputtering yield of 25 molecules per incident He$^+$ ion (mass 4) which is consistent with their yield of $\sim$30 molecules per 3 keV H$_3^+$ ion (mass 3) at higher temperatures (15-20 K) and normal incidence.

During irradiation, we monitored the sputtered material using the QMS. For example, Figure \ref{fig:O2Sputtering} shows the sputtered species observed when O$_2$ ice was irradiated by Ar$^+$ ions. In all cases, we found that O$_2$ is the dominant ejected species, in good agreement with \citet{Ellegaard94,Ellegaard86}. Other species monitored were O (mass 16), O$_3$, (mass 48), and the projectile ion mass. The measured ratio of O$_2$ to O is the same during sputtering as when O$_2$ is fed into the chamber with no beam. This indicates that the O signal arises from the dissociation of O$_2$ in the ioniser of the QMS. Also, the O$_3$ signal with no beam is the minimum detectable pressure in the chamber so it represents the zero level. 

In addition, we performed thermal programmed desorption (TPD) of the O$_2$ ice after irradiation with $^{13}$C$^+$ ions, which consisted of QMS monitoring of the molecular species desorbed from the ice as the temperature increased (Figure \ref{fig:TPD}). We selected to track masses that might be expected to form based on the projectile ion and target ice. The TPD showed no formation of CO or CO$_2$, but we found an O$_3$ peak at temperature $\sim$56 K which is consistent with predictions by \citet{Ellegaard86} and the experimental measurements of  \citet{Ennis}. These authors conducted a comprehensive study of O$_3$ formation in O$_2$ ice at 12 K irradiated by 5 keV singly charged ions of C, N and O. They observed that the O$_3$ yield did not depend on the mass of the projectile ions and concluded that electronic stopping regimes dominate. In a study of O$_2$ ice irradiated by 100 keV protons, \citet{Fama02} find that the formation of O$_3$ quenches the sputtering yield of O$_2$. Our experiments never enter this regime, as evidenced by the linear relation between sputter yield and ion dose (see Figures \ref{fig:SinglySputtering} and \ref{fig:DoublySputtering}).

\section{Conclusions}
\label{S:4}

We have measured the sputtering yield from oxygen ice deposited and irradiated at 10 K by 4 keV singly charged He$^+$, C$^+$, N$^+$, O$^+$ and Ar$^+$ and doubly charged C$^{2+}$, N$^{2+}$ and O$^{2+}$.  Our findings are as follows:

\begin{itemize}

\item The sputtering yields increase linearly with  incident ion mass or with ion momentum squared for all studied species apart from O. The yield varies from $25\pm1$ molecules/ion for He$^+$ (mass 4) to \mbox{$252\pm15$} molecules/ion for Ar$^+$ (mass 40) at a rate of 6.79 O$_2$ ice molecules removed per incoming ion per unit mass of the incident ion. 
\item When using C, N and O ions, the experimental sputtering yields are the same for singly charged and doubly charged ion impact within the experimental errors, as these possess effectively the same momentum.
\item The experimental yields for singly charged ions have been compared with a theoretical model adapted from \citet{Fama} and the results are in good agreement within our experimental errors except in the case of oxygen ions.
\item Preliminary analysis of the ejected species during sputtering with a quadrupole mass spectrometer showed O$_2$ to be the dominant sputtered species. Temperature programmed desorption of the oxygen ice after irradiation with $^{13}$C$^+$ showed the formation of ozone, which is in good agreement with the findings of \citet{Ennis}.

\end{itemize}

\section*{Acknowledgments}

We thank the Leverhulme Trust for support through a Research Project Grant RPG-2013-389. PL is grateful for support from the Royal Society in the form of a Newton Fellowship. Parts of this work were financed by the European Commission's 7th Framework Programme under Grant Agreement No.\ 238258.

\bibliographystyle{model1-num-names}
\bibliography{references}

\end{document}